\documentclass[twocolumn]{revtex4-2}
\usepackage{graphicx}
\usepackage{epstopdf}
\usepackage[rgb]{xcolor}
\definecolor{myblue}{RGB}{0, 0, 255}
\usepackage[colorlinks=true, letterpaper=true, pdfstartview=FitV, linkcolor=myblue, urlcolor=myblue, citecolor=myblue]{hyperref}
\usepackage{amsmath}
\usepackage{amssymb}
\usepackage{float}
\begin{document}

	\title{Magnetic doping-induced second-order and first-order topological phase transition in the photonic alloy}
	\author{Xianbin Wu,$^{1}$ Tiantao Qu,$^{1}$ Xiaoxuan Shi,$^{1}$ Lei Zhang,$^{2,3,*}$ and Jun Chen$^{1,3,\dagger}$}
	\address{$^1$State Key Laboratory of Quantum Optics Technologies and Devices, Institute of Theoretical Physics, Shanxi University, Taiyuan 030006, China\\
		$^2$State Key Laboratory of Quantum Optics Technologies and Devices, Institute of Laser Spectroscopy, Shanxi University, Taiyuan 030006, China\\
		$^3$Collaborative Innovation Center of Extreme Optics, Shanxi University, Taiyuan 030006, China
	}

	\begin{abstract}
The bulk-edge correspondence principle, a cornerstone of topological physics, ensures that first-order topological systems host robust chiral edge states in two dimension. This was later extended to higher-order phases, where second-order topological insulators exhibit localized, topologically protected corner states. While the transition between these distinct phases has been demonstrated in periodic systems, its existence in disordered platforms remains an open question. Here, we demonstrate a controllable topological phase transition between a second-order topological phase and a first-order topological phase in a two-dimensional photonic alloy. By tuning the magnetic doping concentration—implemented by attaching permanent magnets randomly to nonmagnetized yttrium iron garnet rods in an alternately magnetized honeycomb lattice with $C_{3}$ rotational symmetry—we flexibly control the system's topology. At zero doping, we observe higher-order corner states, confirmed by a trivial Chern number and non-zero bulk polarizations of $1/3$. As doping concentration increases, these corner states progressively merge with the bulk states, culminating in the closure of the bulk transmission gap. After the bulk transmission gap reopens with further increased doping, the system transitions to a first-order topological phase, characterized by a nontrivial Chern number of $-1$ and the emergence of a chiral edge state. This transition is reversible, providing a highly tunable and experimentally simple platform for flexibly switching between localized corner states and delocalized chiral edge states within a single photonic system.
    \end{abstract}

	\maketitle

	\section{Introduction}
The bulk-edge correspondence principle first identified in electronic systems \cite{RevModPhys.82.3045,RevModPhys.83.1057,PhysRevLett.61.2015,PhysRevLett.95.226801,PhysRevB.106.195304,PhysRevB.104.094201,zhang2026recentprogressdisorderinducedtopological,PhysRevB.104.L161118} was later observed in photonic platforms \cite{Lu2014,Liu2022,PhysRevLett.125.263603,PhysRevLett.126.067401,PhysRevLett.113.113904,PhysRevLett.115.253901,PhysRevLett.131.213801,doi:10.1126/science.adr5234,Zhou2020,PhysRevLett.132.113802,Wang2023,PhysRevLett.125.133603,PhysRevLett.133.133802,Stützer2018}, demonstrating the broad applicability of topological band theory beyond fermionic systems. Specifically, for a $d$-dimensional ($d$D) topological Chern insulator, the sum of Chern numbers of all bands below a topological band gap is nonzero, and it supports the emergence of ($d-1$)D gapless edge states within the band gap. In real space, these edge modes are confined to the system boundaries and exhibit key features such as unidirectional propagation, immunity to backscattering, and robustness against disorder. The number of edge states is determined by the absolute value of the band gap's Chern number. Systems exhibiting this form of topological protection are typically classified as first-order topological systems. Based on this generalized bulk-edge correspondence principle, several photonic analogs of electronic topological phases have been realized, including the photonic quantum Hall effect \cite{PhysRevLett.100.013904,PhysRevLett.100.013905,Wang2009}, photonic quantum spin Hall effect \cite{PhysRevLett.114.223901,Xu:16,Cheng2016,PhysRevA.95.043827}, and photonic quantum valley Hall effect \cite{Dong2017,doi:10.1126/sciadv.aap8802,PhysRevB.96.020202,Kang2018}. Similar topological phenomena have also been observed in other systems, such as phononic crystals \cite{PhysRevLett.116.093901,Lu2017,Li2025,Yang2023,PhysRevLett.122.014302}, transmission line networks \cite{Jiang2019,15-20200258} and electric circuits \cite{Yang2021,PhysRevLett.114.173902}. The inherent robustness of first order topological systems against defects and disorder has enabled a range of practical implementations, including topological lasers \cite{doi:10.1126/science.aar4003}, on-chip photonic communication systems \cite{Yang2020}, and quantum optical devices \cite{doi:10.1126/science.aaq0327}.
    
In 2017, W. A. Benalcazar, B. A. Bernevig, and T. L. Hughes expanded this field by introducing the concept of bulk-edge-corner correspondence. They formulated a theory of quantized multipole moments in crystalline insulators, extending conventional bulk-edge correspondence to a more general framework \cite{doi:10.1126/science.aah6442,PhysRevB.96.245115}. Within this framework, a $2$D second-order topological insulator possesses $1$D gapped edge states as well as $0$D in-gap corner states. These corner states, typically localized at system corners, are protected by nontrivial topology characterized by quantized Wannier centers \cite{Li2020} or quadrupole moments \cite{He2020}. Following this theoretical development, second-order topological insulators have been realized in various platforms, including photonic systems \cite{PhysRevB.107.014105,He2020,PhysRevB.98.205147,PhysRevLett.132.073803,PhysRevLett.122.233903,10.1093/nsr/nwae121,PhysRevResearch.2.042038,PhysRevLett.134.116607,Li2020}, acoustic systems \cite{Xue2019,Ni2019,10.1063/5.0193895}, and electronic systems \cite{PhysRevB.106.064109,PhysRevLett.120.026801,PhysRevB.101.195309,Long2023}.

A compelling question therefore arises: can a transition occur between second-order topological phase (SOTP) and first-order topological phase (FOTP)? Research on this question remains limited. A recent research has shown that transitions between SOTP and FOTP can be induced by adjusting the ratio of magnetized yttrium iron garnet (YIG) rod radius to lattice constant, or by altering external magnetic field strength in periodic photonic systems \cite{10.1093/nsr/nwae121}. But could such transitions also exist in non-periodic photonic systems? Topological photonic alloys \cite{PhysRevLett.132.223802,PhysRevB.110.094206,doi:10.1021/acsphotonics.4c01412,10.1063/5.0232244,gw3w-5h94} provide simple yet effective disordered platforms for achieving topological phase transitions, enabling transitions from topologically trivial to nontrivial states through random mixing of nonmagnetized and magnetized YIG rods in non-periodic two-dimensional photonic configurations. Using the alloy approach, could we achieve a transition between SOTP and FOTP in disordered photonic systems?

\begin{figure}[htbp]
	   \centering
    \includegraphics[width=1\columnwidth]{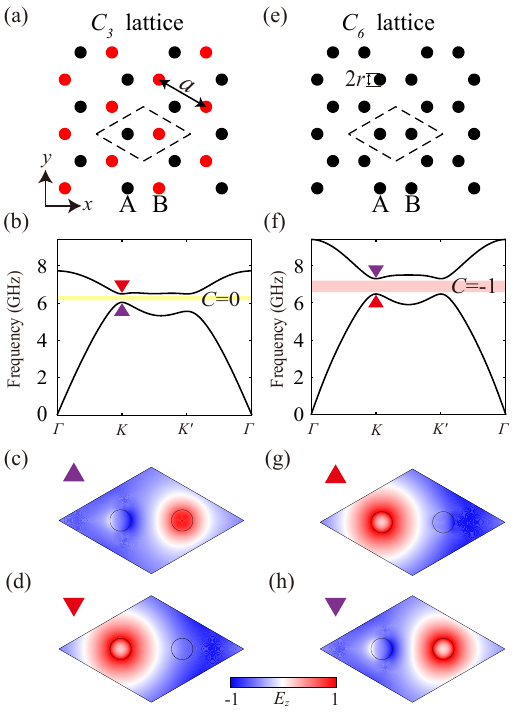} 
	   \caption{Band structures of honeycomb lattice photonic crystals with two different symmetries: $C_{3}$ rotational symmetry ($C_{3}$ lattice, with YIG rods alternately magnetized) and $C_{6}$ rotational symmetry ($C_{6}$ lattice, with all YIG rods magnetized). (a), (e) Schematics of the $C_{3}$ and $C_{6}$ lattices, respectively. Black (red) circles denote magnetized (nonmagnetized) YIG rods. The dashed rhombi represent the primitive cells of the two lattices, and sites $A$ and $B$ correspond to two distinct atoms in the primitive cells. $a$ represents the lattice constant and $r$ is the radius of the YIG rods. (b), (f) Band structures of the $C_{3}$ and $C_{6}$ lattices, respectively. The yellow and pink shaded regions denote the first photonic band gaps of the $C_{3}$ and $C_{6}$ lattices, with their associated Chern numbers labeled. (c)-(d) Normalized $E_{{z}}$ field distributions of the eigenmodes at the $K$ point for the first and second bands of the $C_{3}$ lattice, respectively. These eigenmodes are indicated by the purple upward and red downward triangles in panel (b). (g)-(h) Normalized $E_{{z}}$ field distributions of the eigenmodes at the $K$ point for the first and second bands of the $C_{6}$ lattice, marked by the red upward and purple downward triangles in panel (f). In both symmetry configurations, the lattice constant is set to $a = 20\ mm$ and the radius of each YIG rods is $r = 2\ mm$.}\label{fig:1}
\end{figure}
In this paper, we demonstrate such transitions, successfully converting second-order topological corner states (SOTP) to a first-order topological chiral edge state (FOTP) and vice versa by adjusting magnetic doping concentration in a hexagonal photonic alloy. The tuning of the magnetic doping concentration can be achieved in honeycomb lattice photonic crystals with $C_{3}$ rotational symmetry (with alternately magnetized YIG rods) by simply attaching permanent magnets randomly to nonmagnetized YIG rods. The topological transition is marked by the closure and reopening of the bulk transmission gap (the first band gap). At zero magnetic doping ($C_{3}$ lattice), the band below the gap has a Chern number of $0$ and bulk polarizations $P_{1} = P_{2} = 1/3$. This indicates the absence of gapless chiral edge states but confirms the presence of higher-order topological corner states protected by quantized Wannier centers. As doping concentration increases, topological corner states in the bulk transmission gap progressively merge into extended bulk states, accompanied by the gap's closure. With further increased doping, the bulk transmission gap reopens. Using the scattering method, a Chern number of $-1$ is obtained, confirming the emergence of a topological chiral edge state. This verifies a topological phase transition from SOTP to FOTP accompanying the bulk transmission gap's closure and reopening. This transition can also be reversed—from FOTP back to SOTP—by reducing magnetic doping concentration (simply removing permanent magnets randomly from YIG rod ends).
\begin{figure*}[htbp]
       \centering 
  \includegraphics[width=1\textwidth]{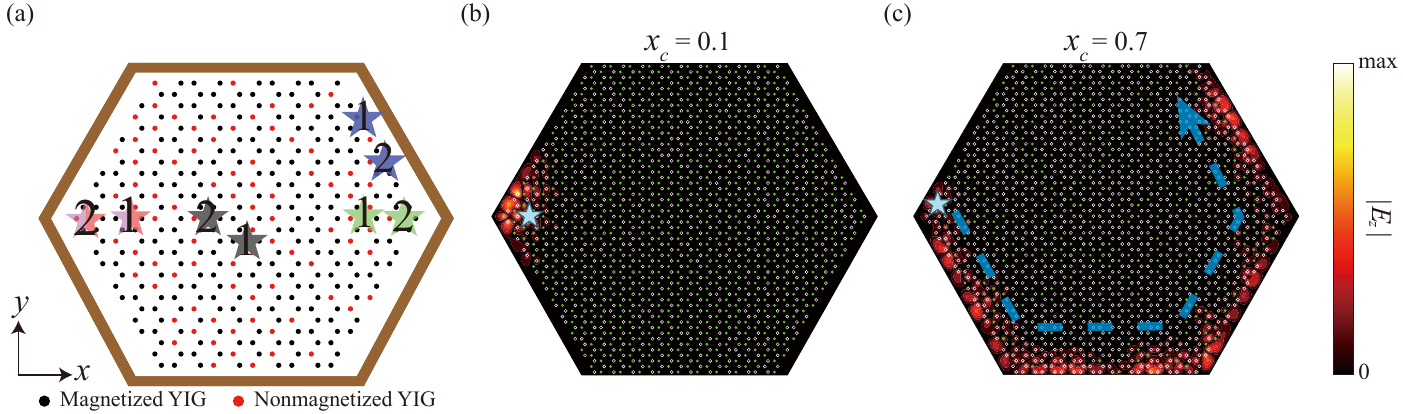} 
       \caption{Topologically protected localized corner states and the delocalized chiral edge state realized within a photonic alloy. (a) Schematic of the photonic alloy structure, constructed by randomly replacing some nonmagnetized YIG rods (red dots) with magnetized YIG rods (black dots) in a $2$D alternately magnetized honeycomb lattice photonic crystal with $C_{3}$ rotational symmetry [see Fig. \ref{fig:1}(a)]. The brown boundary represents the perfect electric conductor (PEC). Colored stars labeled with numbers $1$ and $2$ denote the positions of the source and probe antennas used to obtain corner, edge, and bulk transmissions, respectively. (b)-(c) Normalized electric field distributions $\left | E_{{z}}  \right |$ under line sources excitation for doping concentrations $x_{c} = 0.1$ and $x_{c} = 0.7$, respectively. Hollow white circles represent magnetized YIG rods, green dots represent nonmagnetized YIG rods, the blue stars indicate the locations of the line sources, and the blue arrow shows the propagation direction of the chiral edge state. The upper boundary in (c) uses an absorbing boundary condition, while the remaining boundaries are PEC, the same as in (a).}\label{fig:2}
\end{figure*}

	\section{Model and results}
In the construction of $2$D photonic alloys, YIG (a gyromagnetic material) is used for the rods. Permanent magnets can be attached to the rod ends to magnetize them individually \cite{PhysRevLett.132.223802}, causing them to exhibit gyromagnetic behavior in the microwave regime. Specifically, the YIG rods have a relative permittivity of 15.26, and their relative permeability tensor takes the following form \cite{pozar2021microwave}:
	\begin{equation}
          \tilde{\mu} = \begin{pmatrix}  
          \mu _{r} & i\kappa  & 0 \\  
          -i\kappa & \mu _{r} & 0 \\  
          0 & 0 & 1  
          \end{pmatrix} ,
	\end{equation}
where $\mu _{r} = 1+\frac{\omega _{m}\left ( \omega _{0}+i\alpha \omega  \right )}{\left ( \omega _{0}+i\alpha \omega  \right )^{2}-\omega ^{2}}$ and $\kappa = \frac{\omega _{m}\omega }{\left ( \omega _{0}+i\alpha \omega  \right )^{2}-\omega ^{2}}$. $\omega$ is the working frequency, the characteristic frequency $\omega _{m} = 4\pi \gamma M_{s}$, the resonance frequency $\omega _{0} = \gamma H_{0}$, the gyromagnetic ratio $\gamma = 2.8$ MHz/Oe, the saturation magnetization $4\pi M_{s} = 1884$ Oe,  the external magnetic field $H_{0} = 1150$ Oe is applied along the z direction, and $\alpha$ represents the damping coefficient, which we set to $0$ in our numerical simulations. YIG material exhibits low intrinsic losses \cite{PhysRevB.107.014105}, which have a negligible impact on our primary findings. Therefore, losses are omitted here. For a detailed comparative analysis of results incorporating realistic YIG losses, see Sec. I of the Supplemental Material \cite{supp}.

How can we design photonic alloys to enable the transition between SOTP and FOTP? We gain insights from examining the band structures of gyromagnetic honeycomb lattice photonic crystals with different symmetries: $C_{6}$ rotational symmetry (where all YIG rods are magnetized) \cite{PhysRevLett.106.093903} and $C_{3}$ rotational symmetry (where YIG rods are alternately magnetized) \cite{PhysRevB.107.014105}, as illustrated in Fig. \ref{fig:1}. The band structures for the transverse magnetic (TM) modes of both the $C_{3}$ and $C_{6}$ lattices are shown in Figs. \ref{fig:1}(b) and \ref{fig:1}(f), respectively. For the $C_{3}$ lattice, a photonic band gap opens between the first and second bands, spanning the frequency range from $6.04$ to $6.49$ GHz. Using the Wilson-loop approach \cite{Wang_2019}, we verify that the Chern number of the first band is $0$, while the bulk polarizations are $P_{1} = P_{2} = 1/3$ (details provided in the APPENDIX \ref{APPENDIX}). This confirms the absence of gapless chiral edge states in this gap, but indicates the presence of higher-order topological corner states protected by quantized Wannier centers. The corresponding eigenmode field distributions at the $K$ point for the first and second bands are shown in Figs. \ref{fig:1}(c) and \ref{fig:1}(d), associated with the purple and red triangles in Fig. \ref{fig:1}(b), respectively. In contrast, the $C_{6}$ lattice exhibits a band gap between the lowest two bands, ranging from $6.48$ to $7.30$ GHz. The Chern number of the lowest band is computed to be $-1$, indicating the existence of a gapless chiral edge state. The eigenmode electric field distributions at the $K$ point for these two bands are displayed in Figs. \ref{fig:1}(g) and \ref{fig:1}(h), corresponding to the red and purple triangles Fig. \ref{fig:1}(f). Thus, the band gaps of the $C_{3}$ and $C_{6}$ lattices support distinct topological phases: the $C_{3}$ lattice exhibits SOTP, while the $C_{6}$ lattice displays FOTP. Furthermore, their eigenmode distributions at the K point show a feature reminiscent of band inversion, indicating the existence of a critical point associated with a topological phase transition. Therefore, the photonic alloy used to achieve the transition between SOTP and FOTP will be based on these lattices through random doping.
 
Specifically, we construct a photonic alloy by randomly replacing some nonmagnetized YIG rods with magnetized YIG rods in a $2$D alternately magnetized honeycomb lattice photonic crystal with $C_{3}$ rotational symmetry, as shown in Fig. \ref{fig:2}(a). We define the doping concentration as $x_{c} = (N_{A}-N_{B})/(N_{A}+N_{B})$, where $N_{A}$ and $N_{B}$ represent the numbers of magnetized and nonmagnetized YIG rods, respectively. When $N_{B} = N_{A}$, the doping concentration $x_{c} = 0$, and the system remains in its initial state ($C_{3}$ lattice). When $N_{B} = 0$, the doping concentration reaches $x_{c} = 1$, meaning all nonmagnetized YIG rods have been replaced by magnetized ones. It is found that a photonic alloy with a low doping concentration ($x_{c} = 0.1$) supports a localized corner state, as shown in Fig. \ref{fig:2}(b). This corner state can be excited near the left corner of the hexagonal structure using a line source at $6.35$ GHz (eigenfrequency of this corner state). In contrast, a photonic alloy with a higher doping concentration ($x_{c} = 0.7$) exhibits a chiral edge state when excited at $6.56$ GHz, as shown in Fig. \ref{fig:2}(c). Although random magnetic doping breaks exact crystalline symmetries in photonic alloys, the robustness of topological modes is protected by the mobility gap \cite{PhysRevB.104.094201}, real-space topological invariants \cite{PhysRevB.104.094201,PhysRevLett.129.043902}, and ``average symmetry" in the thermodynamic limit \cite{PhysRevB.106.195304}. A detailed analysis of these protection mechanisms and the robustness of the corner state and chiral edge state in Figs. \ref{fig:2}(b)–2(c) is provided in Sec. II of the Supplemental Material \cite{supp}. These results in Fig. \ref{fig:2} demonstrate that both topologically protected chiral edge transport and localized corner states can be realized within a single disordered photonic platform. This capability enables the integration of FOTP and SOTP, offering a promising approach for robust, tunable switching between delocalized and localized photonic states.

\begin{figure}[htbp]
	   \centering
    \includegraphics[width=1\columnwidth]{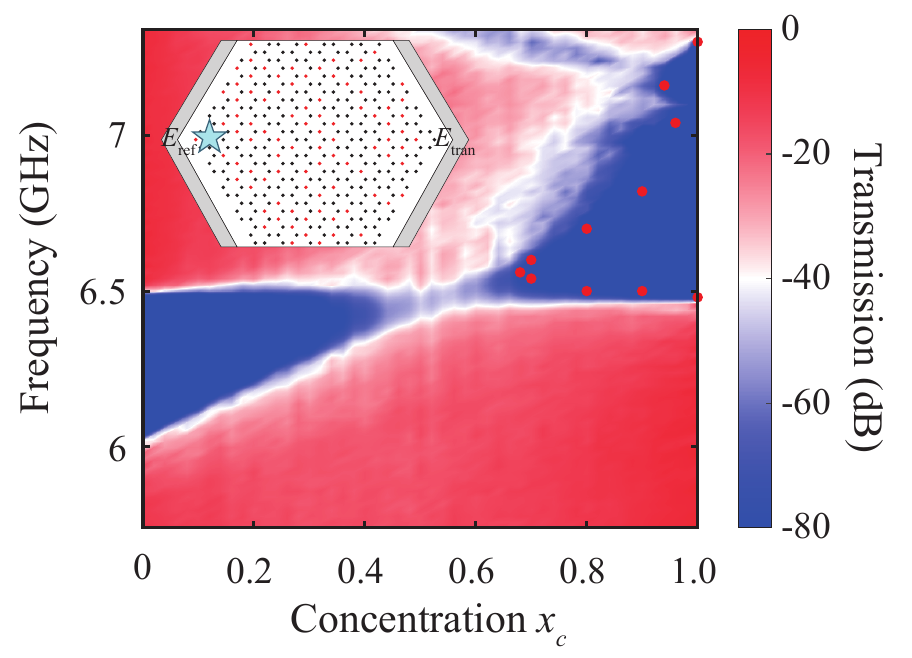} 
	   \caption{Bulk transmission spectrum as a function of doping concentration $x_{c}$. Each data point represents the average of twenty disorder samples. The red dots in the high-doping region mark the boundary of the topological gap characterized by a Chern number of $-1$, as determined using the scattering method. Inset: Schematic of numerically calculated transmission in the photonic alloy system. The top and bottom boundaries are continuously connected, while the left and right boundaries are enclosed using absorbing boundary conditions. The location of the line source is marked by a blue star.}\label{fig:3}
   \end{figure}

To further confirm the existence of the topological phase transition, we compute and analyze the bulk transmission spectrum of the photonic alloy system at various doping concentrations, as presented in Fig. \ref{fig:3}. This method is widely used for identifying topological phase transitions in disordered systems \cite{PhysRevLett.132.223802,PhysRevLett.129.043902,PhysRevLett.133.133802}. For bulk transmission calculations, we connect the top and bottom of the photonic alloy with continuous periodic boundary conditions, while applying absorbing boundary conditions to the left and right sides. A line source is placed near the left boundary, as shown in the inset of Fig. \ref{fig:3}. The time-averaged Poynting vector is integrated over the left and right boundaries to calculate the energy. We define $E_{tran}$ as the energy passing through the photonic alloy (via the right boundary) and $E_{ref}$ as the energy directly reflected by the photonic alloy (exiting through the left boundary). The total energy leaving the system is thus $E_{tot} = E_{tran} +E_{ref}$, and the bulk transmission is given by \cite{PhysRevLett.115.253901} 
 	\begin{equation}
     \left \langle T \right \rangle = 20\left \langle\log_{10}\left (  {\frac{E_{tran} }{E_{tot} } }  \right )   \right \rangle .
	\end{equation} 
In Fig \ref{fig:3}, red regions denote bulk states that allow energy to exit through the right boundary, while blue regions (the transmission gaps) correspond to near-zero transmission caused by corner states or edge states. As the doping concentration $x_{c}$ increases, the bulk transmission gap closes and then reopens. The gap appears not to close completely, primarily due to momentum mismatch and limited coupling efficiency of the line source (see detailed explanations in Sec. III of the Supplemental Material \cite{supp}). At the leftmost and rightmost parts of the bulk transmission spectrum, the $C_{3}$ lattice ($x_{c} = 0$) supports higher-order topological corner states, while the $C_{6}$ lattice ($x_{c} = 1$) supports chiral edge states.  Furthermore, the robustness of these topological modes is protected by the transmission gap (mobility gap \cite{PhysRevLett.129.043902}) \cite{PhysRevB.104.094201}. As long as this gap remains open, corner or edge modes are shielded from disorder. Consequently, the distinct topological properties before and after the closure of the bulk transmission gap confirm that a phase transition has occurred (see detailed discussions regarding the phase transition between SOTP and FOTP in Sec. III of the Supplemental Material \cite{supp}). How, then, does the topological phase gradually evolve as the doping concentration $x_{c}$ increases?

\begin{figure*}[htbp]
	   \centering
    \includegraphics[width=1\textwidth]{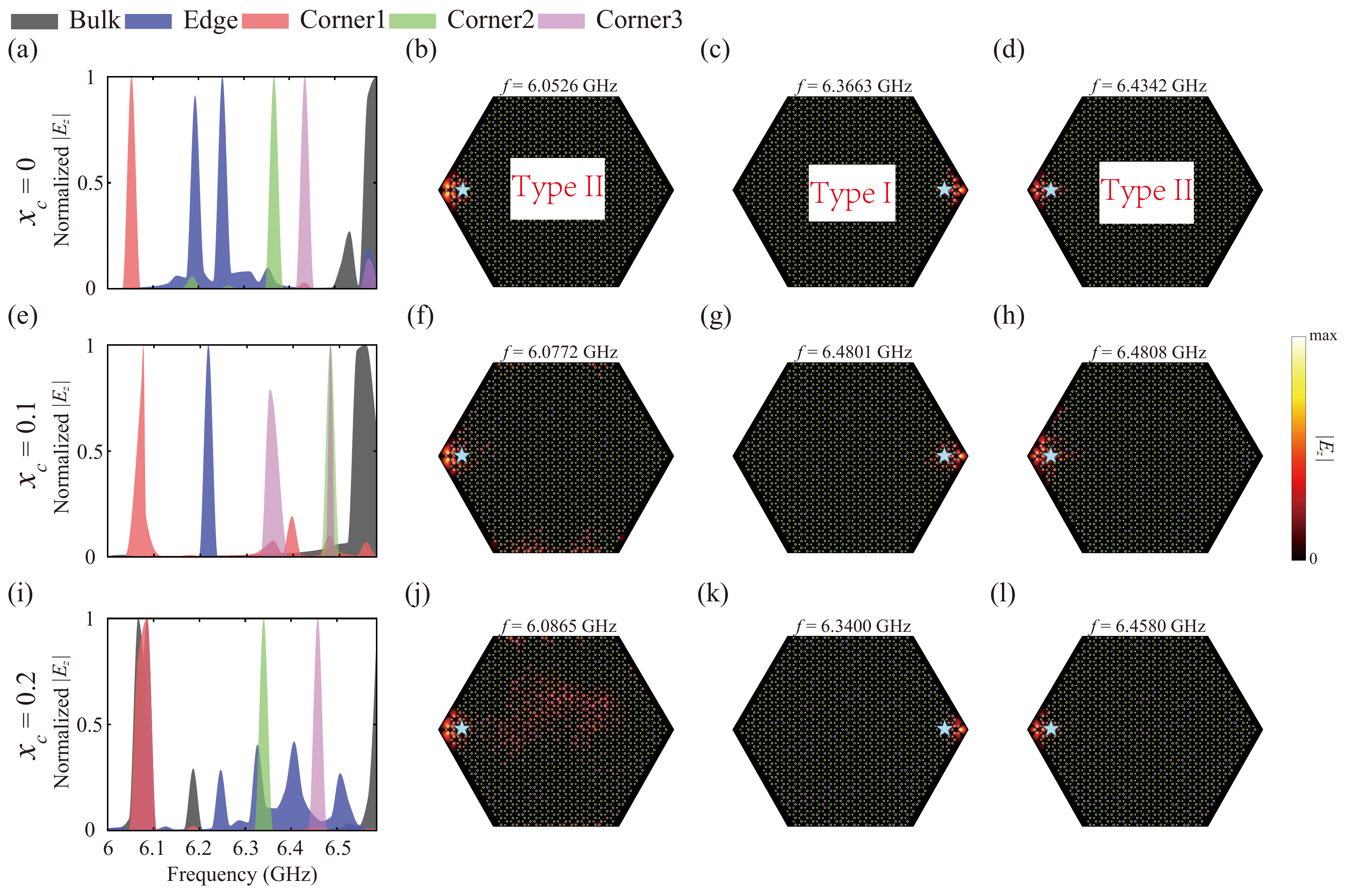} 
	   \caption{Frequency-resolved transmissions and field distributions for the photonic alloy. (a), (e), (i) Frequency-resolved transmissions of corner, edge, and bulk states at doping concentrations $x_{c} =$ $0$, $0.1$, and $0.2$, respectively. The colors of the transmission peaks correspond to the colors of the source-probe pairs used in Fig. \ref{fig:2}(a) to excite and collect information about these states. (b)-(d) Distributions of the absolute electric field value ($\left | E_{{z}}  \right |$) for the three topological corner states excited  in order of increasing frequency when the photonic alloy's doping concentration $x_{c} = 0$ ($C_{3}$ lattice). (f)-(h), (j)-(l) Field distributions showing the localization of corner states that evolved from the three topological corner states in (b)-(d) at doping concentrations $x_{c} = 0.1$ and $x_{c} = 0.2$. Blue stars mark the line source locations in (b)-(d), (f)-(h), and (j)-(l). Other simulation parameters are identical to those used in Figs. \ref{fig:2}(b)-\ref{fig:2}(c).}\label{fig:4}
\end{figure*}

To characterize the topological phase evolution of the photonic alloy before phase transition, we investigate the frequency-resolved transmissions of corner, edge, and bulk states at varying doping concentrations, as shown in Fig. \ref{fig:4}. Without loss of generality, we randomly use one sample for calculations at each doping concentration. As doping concentration increases, the evolution of topological corner states follows a consistent trend across different random samples. See detailed discussions in Sec. IV of the Supplemental Material \cite{supp}. We place a line source (labeled as `1') at representative locations in the photonic alloy schematiced in Fig. \ref{fig:2}(a), then collect the excited electric field at the probe (labeled as `2'). In Fig. \ref{fig:2}(a), the source-probe pair for bulk transmission is marked with gray pentagrams, edge transmission with blue pentagrams, and corner transmission with red, green, and purple pentagrams. These colors correspond to the frequency-resolved transmissions shown in Figs. \ref{fig:4}(a), \ref{fig:4}(e), and \ref{fig:4}(i). We use three different colors to describe corner transmissions because the $C_{3}$ lattice (at doping concentration $x_{c} = 0$) exhibits two types of topological corner states: type I [Fig. \ref{fig:4}(c)] and type II [Figs. \ref{fig:4}(b) and \ref{fig:4}(d)]. The distinction between type-I and type-II corners depends on whether the corner rod has been applied a magnetic field \cite{PhysRevB.107.014105}. Additionally, with the angular bisector as the axis of symmetry and based on the electric field $E_{{z}}$ distribution (magnitude and polarity), the type-II corners include two modes: an antisymmetric mode at $6.0526$ GHz [Fig. \ref{fig:4}(b)] and a symmetric mode at $6.4342$ GHz [Fig. \ref{fig:4}(d)].

Figure \ref{fig:4}(a) shows that at doping concentration $x_{c} = 0$, the system exhibits three transmission peaks around $6.0526$ GHz (red peak), $6.3663$ GHz (green peak), and $6.4342$ GHz (purple peak), corresponding to topological corner states (SOTP). The system can also excite topologically trivial edge states \cite{PhysRevB.107.014105} (blue peaks) and bulk states at high frequencies (gray peaks). As the doping concentration increases to $x_{c} = 0.1$, Fig. \ref{fig:4}(e) shows that the corner transmission peaks persist. Compared to Fig. \ref{fig:4}(a), the frequencies of the corner states corresponding to the red and green peaks in Fig. \ref{fig:4}(e) have shifted. Their  electric field distribution characteristics, shown in Figs. \ref{fig:4}(f) and \ref{fig:4}(g), indicate that these two localized corner states have evolved from the topological corner states shown in Figs. \ref{fig:4}(b) and \ref{fig:4}(c). Interestingly, the topological corner state corresponding to the purple peak in Fig. \ref{fig:4}(a) [field distribution in Fig. \ref{fig:4}(d)] splits into two purple peaks at doping concentration $x_{c} = 0.1$, as shown in Fig. \ref{fig:4}(e). These peaks correspond to frequencies of $6.3500$ GHz [field distribution shown in Fig. \ref{fig:2}(b)] and $6.4808$ GHz [field distribution shown in Fig. \ref{fig:4}(h)], with the latter occurring at an excitation frequency very close to the corner state of the green peak. As the doping concentration further increases to $x_{c} = 0.2$, a notable phenomenon occurs: while the localized corner states evolved from the two higher-frequency topological corner states [green and purple peaks in Fig. \ref{fig:4}(a), with corresponding field distributions in Figs. \ref{fig:4}(c) and \ref{fig:4}(d)] still exist independently [green and purple peaks in Fig. \ref{fig:4}(i), with field distributions in Figs. \ref{fig:4}(k) and \ref{fig:4}(l)], the corner state [red peak in Fig. \ref{fig:4}(i) with field distribution in Fig. \ref{fig:4}(j)] evolved from the lower-frequency topological corner state [red peak in Fig. \ref{fig:4}(a) with field distribution in Fig. \ref{fig:4}(b)] has begun to merge with a nearby bulk state. This merger is visible in the field distribution shown in Fig. \ref{fig:4}(j). This indicates that the lower-frequency localized corner state gradually transforms into an extended bulk state. Bulk states emerging in this frequency range correspond with Fig. \ref{fig:3}'s bulk transmission spectrum before the phase transition, where the lower band's upper edge rises as doping concentration increases. Further calculations at higher doping concentrations reveal that corner states do not necessarily merge in sequence from low-frequency to high-frequency states with the lower bulk band. Higher-frequency corner states may merge into the upper bulk band, causing them to merge with bulk states before intermediate-frequency corner states (see additional numerical results and discussions in Sec. V of the Supplemental Material \cite{supp}). Our analysis suggests that as doping concentration increases, all topologically protected localized corner states will eventually merge into extended bulk states. This conclusion can be understood as follows: At the leftmost part of the bulk transmission spectrum shown in Fig. \ref{fig:3}, the $C_{3}$ lattice ($x_{c} = 0$) supports higher-order topological corner states. Since the robustness of topological modes is protected by the transmission gap, the localized corner states persist despite disorder within the transmission gap on the left side of the bulk transmission spectrum. However, as doping concentration increases, the transmission gap gradually shrinks and closes, meaning the topological protection is ultimately lost. Consequently, all corner states will inevitably merge into extended bulk states. Furthermore, our results show that doping concentration can effectively tune the excitation frequency of corner states in the photonic alloy. More intriguingly, corner states can either split or merge with bulk states depending on this concentration, offering a flexible mechanism for controlling optical field localization in photonic systems.

\begin{figure}[htbp]
	   \centering
    \includegraphics[width=1\columnwidth]{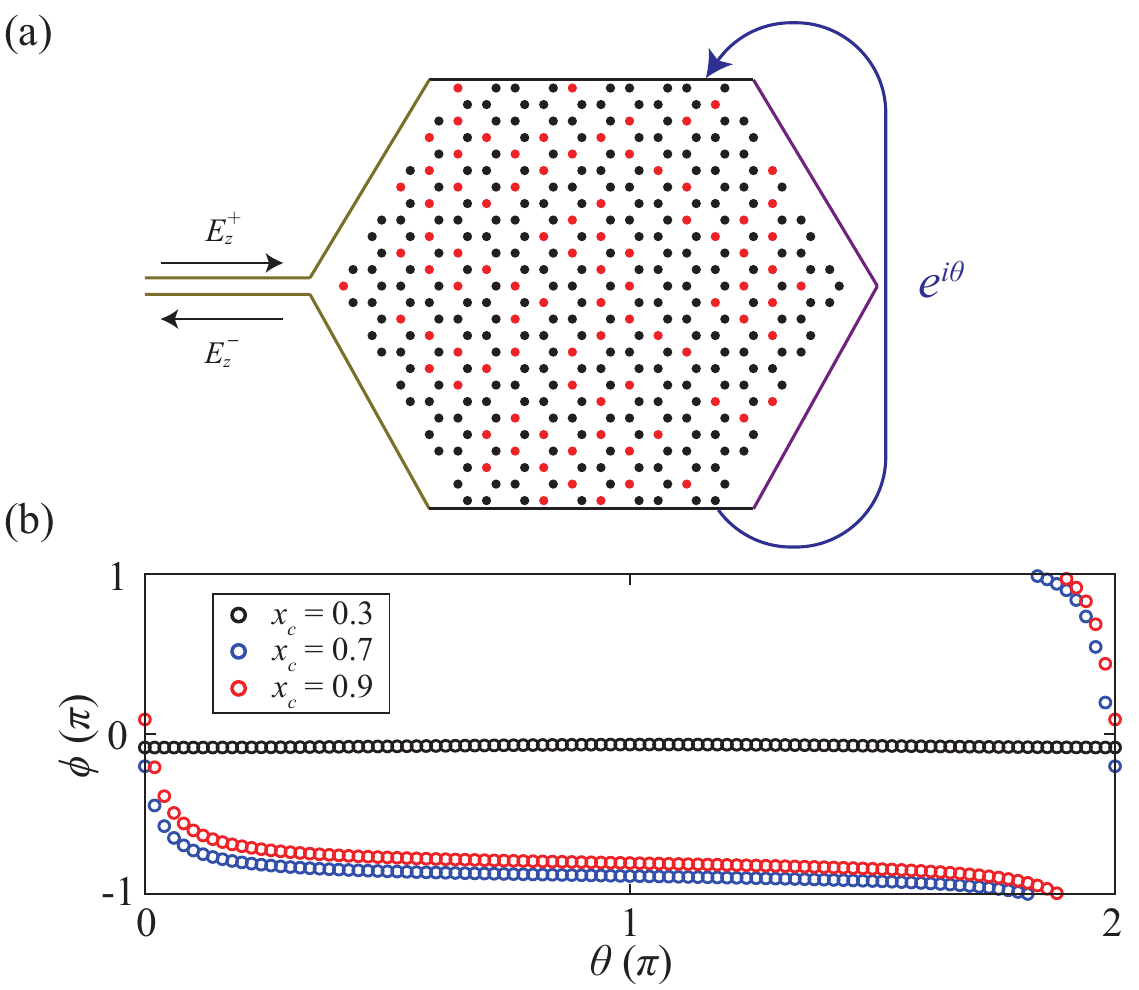} 
	   \caption{Characterization of topological behavior in photonic alloys using the scattering method. (a) Schematic of the setup for extracting topological signatures from a hexagonal photonic alloy. The photonic alloy has a twisted boundary condition $\psi$ (\textit{y} = L) =$\psi$ (\textit{y} = 0)$e^{i\theta }$ applied to its top ($\textit{y} = L$) and bottom ($\textit{y} = 0$) boundaries. The left side connects to an air lead bounded by a perfect magnetic conductor, while the right boundary has an absorbing condition. A TM-polarized incident wave $E_{{z}}^{+}$ enters the system, and the reflected waveguide modes $E_{{z}}^{-}$ are analyzed. (b) Reflection phase $\phi $ plotted against twisting angle $\theta $ for different doping concentrations. Black, blue, and red hollow circles represent doping concentrations of $x_{c} =$ $0.3$, $0.7$, and $0.9$, respectively.}\label{fig:5}
\end{figure}

As the doping concentration further increases, the bulk transmission gap shown in Fig. \ref{fig:3} reopens. To accurately characterize the topological behavior of the photonic alloy after the phase transition, we employ the scattering method \cite{PhysRevLett.129.043902,zijderveld2024scatteringtheoryhigherorder,PhysRevB.85.165409,PhysRevB.83.155429,doi:10.1126/sciadv.adg3186}. As shown schematically in Fig. \ref{fig:5}(a), we impose a twisted boundary condition $\psi$ (\textit{y} = L) =$\psi$ (\textit{y} = 0)$e^{i\theta }$ to the top ($\textit{y} = L$) and bottom ($\textit{y} = 0$) boundaries of the photonic alloy. The left side of the alloy, bounded by a perfect magnetic conductor, connects to an air waveguide lead supporting a single TM mode at the relevant frequency range. The right boundary has an absorbing boundary condition. Similar to the Wilson-loop formalism, where the winding number of the Berry phase along one reciprocal lattice basis vector $k_{i}$ corresponds to the Chern number in periodic systems \cite{Wang_2019}, in disordered photonic systems, the Chern number equals the winding number of the reflection phase $\phi \left ( \theta  \right ) $ (the phase difference between the incident and the reflected waveguide modes) as the twisting angle $\theta$ varies from $0$ to $2\pi $. To examine the topological transport behavior in the reopened gap after the phase transition shown in Fig. \ref{fig:3}, we randomly select an excitation frequency of $f = 6.56$ GHz for doping concentrations $x_{c} = 0.7$ and $x_{c} = 0.9$. For comparison, we have also analyzed the reflection phase winding at $x_{c} = 0.3$ with an excitation frequency of $6.40$ GHz before the topological phase transition. Figure \ref{fig:5}(b) reveals that for $x_{c} = 0.3$, the reflection phase $\phi \left ( \theta  \right ) $ remains nearly constant as $\theta$ changes, indicating a trivial Chern number of $0$. In contrast, for $x_{c} = 0.7$ and $x_{c} = 0.9$, $\phi \left ( \theta  \right ) $ exhibits a full $2\pi $ winding, signaling a nontrivial Chern number of $-1$, which corresponds to the emergence of a topological chiral edge state as shown in Fig. \ref{fig:2}(c). The three lines in Fig. \ref{fig:5}(b) are calculated using a randomly selected sample at each of three different doping concentrations. However, it should be noted that at any given doping concentration, the winding number of the reflection phase is independent of the random sampling of magnetic doping, see more details in Sec. VI of the Supplemental Material \cite{supp}. Using scattering methods, we further calculate the boundary of the topological gap within which all random samples yield a Chern number of -1, marked with red dots in Fig. \ref{fig:3}. It is revealed that the critical doping concentration for the topological chiral edge state is approximately $x_{c} = 0.68$, at which point the photonic alloy transitions to supporting FOTP.

Therefore, increasing the magnetic doping concentration in hexagonal photonic alloys enables a topological transition from second-order topological corner states (SOTP) to the first-order topological chiral edge state (FOTP). Conversely, reducing this magnetic doping concentration—achieved experimentally by simply removing permanent magnets randomly from YIG rod ends—facilitates a transition from FOTP back to SOTP. This approach provides a convenient method for flexibly controlling optical field localization properties within a single photonic system.

    \section{Conclusion and discussion}
In this study, we demonstrated the topological transition between second-order topological corner states (SOTP) to the first-order topological chiral edge state (FOTP) by adjusting magnetic doping concentration in a hexagonal photonic alloy. This adjustment can be implemented in honeycomb lattice photonic crystals with $C_{3}$ rotational symmetry (with alternately magnetized YIG rods) by simply attaching permanent magnets randomly to nonmagnetized YIG rods. In the absence of magnetic doping ($C_{3}$ lattice), the topological properties of corner states in the bulk transmission gap (the first band gap) are verified using the Wilson-loop approach. The first band has a Chern number of $0$ and bulk polarizations $P_{1} = P_{2} = 1/3$. This confirms the absence of gapless chiral edge states while indicating the presence of higher-order topological corner states protected by quantized Wannier centers. Before the bulk transmission gap closes, we investigated the evolution of these topological localized corner states as magnetic doping concentration increases by examining frequency-resolved transmissions of corner, edge, and bulk states. At low magnetic doping concentrations, the photonic alloy supports SOTP. As doping concentration increases, the topological corner states in the bulk transmission gap may split. More importantly, the lowest-frequency topological corner state gradually merges with bulk states, transforming from a localized corner state into an extended bulk state. These emerging bulk states correspond to the bulk transmission spectrum where the lower band's upper edge rises with increasing doping concentration. With further increases in doping concentration, the higher-frequency localized corner states within the bulk transmission gap progressively merge into extended bulk states, coinciding with the bulk transmission gap's closure. As doping concentration increases even further, the bulk transmission gap reopens. Using the scattering method to characterize the topological behavior, our calculations yield a nontrivial Chern number of $-1$,  corresponding to the emergence of a topological chiral edge state. This confirms a topological phase transition from SOTP to FOTP accompanying the closure and reopening of the bulk transmission gap. We can also reverse this process—transitioning from FOTP back to SOTP—by reducing magnetic doping concentration (simply removing permanent magnets randomly from YIG rod ends). This approach requires only a simple lattice structure and straightforward experimental implementation, offering a viable, tunable platform for controlling topological phase transitions between SOTP and FOTP, and providing a convenient method for flexibly controlling optical field localization within a single photonic system. Since this topological phase transition between SOTP and FOTP only requires that the bandgaps of two lattice configurations (the two extremes of random doping) support SOTP and FOTP, respectively, and a band inversion between the corresponding bands, similar phase transition mechanisms and ideas can be extended to acoustic crystals and other systems (see more detailed discussions in Sec. VII of the Supplemental Material \cite{supp}).

\begin{acknowledgments}
We gratefully acknowledge the support from the National Natural Science Foundation of China (Grants No. $12574340$, No. $12474047$, No. $12174231$), the Fund for Shanxi $1331$ Project, research project supported by Shanxi Scholarship Council of China. This research was partially conducted using the High Performance Computer of Shanxi University.	
\end{acknowledgments}

\section*{DATA\ AVAILABILITY}
The data that support the findings of this article are not publicly available. The data are available from the authors upon reasonable request.

\appendix*

\renewcommand{\appendixname}{APPENDIX}

\section{\MakeUppercase{the topological invariants for periodic photonic crystals}}
\label{APPENDIX}
We use $COMSOL\ Multiphysics$ based on finite-element methods to numerically solve Maxwell's equations for determining the band structure of photonic crystals. This approach is mathematically equivalent to solving an eigenvalue problem with a Hermitian operator in the absence of material loss \cite{03c05e45-4061-3158-a571-ca2f2d6dfda7}. The resulting eigenfields correspond to the electromagnetic fields. For periodic photonic crystals, the electric field can be represented as
 	\begin{equation}
    \textbf{E}_{n} \left ( \textbf{\textit{k}} , \textbf{\textit{r}}\right ) = \textbf{u}_{n} \left ( \textbf{\textit{k}} , \textbf{\textit{r}}\right )e ^{i \textbf{\textit{k}} \cdot  \textbf{\textit{r}} } ,
	\end{equation}    
where ${\textbf{u}_{n}} \left ( \textbf{\textit{k}} , \textbf{\textit{r}}\right )$ is the periodic part of the eigenstate at the reciprocal lattice vector $\textbf{\textit{k}}$ for the $n$th band. For a $2$D system, the Chern number of the $n$th band can then be computed by integrating the Berry curvature over the first Brillouin zone \cite{doi:10.1143/JPSJ.74.1674}
	\begin{equation}\label{A}
C_{n} = \frac{1}{2\pi }  \oint_{BZ} \Omega _{n}^{z}\left ( \textbf{\textit{k}} \right )d^{2}\textbf{\textit{k}}. 
	\end{equation}
Here, $\Omega _{n}^{z}$ is the $z$-component of the Berry curvature $\boldsymbol{\Omega}_{n}$, which is obtained as the curl of the Berry connection $\textbf{\textit{A}}_{n}$	
	\begin{equation}
\boldsymbol{\Omega}_{n}\left ( \textbf{\textit{k}} \right ) = \boldsymbol{\nabla}_{\textbf{\textit{k}} }\times \textbf{\textit{A}}_{n}\left ( \textbf{\textit{k}} \right ), 
	\end{equation}
and
	\begin{equation}
    \textbf{\textit{A}}_{n}\left ( \textbf{\textit{k}} \right ) = i\left \langle\textbf {u}_{n} \left ( \textbf{\textit{k}} , \textbf{\textit{r}}\right ) \right | \boldsymbol{\nabla}_{\textbf{\textit{k}} } \left |\textbf{u}_{n} \left ( \textbf{\textit{k}}, \textbf{\textit{r}}\right )  \right \rangle . 
	\end{equation}
Equation (\ref{A}) can be numerically solved through the Wilson-loop approach \cite{Wang_2019}
 \begin{figure}[htbp]
	   \centering
    \includegraphics[width=1\columnwidth]{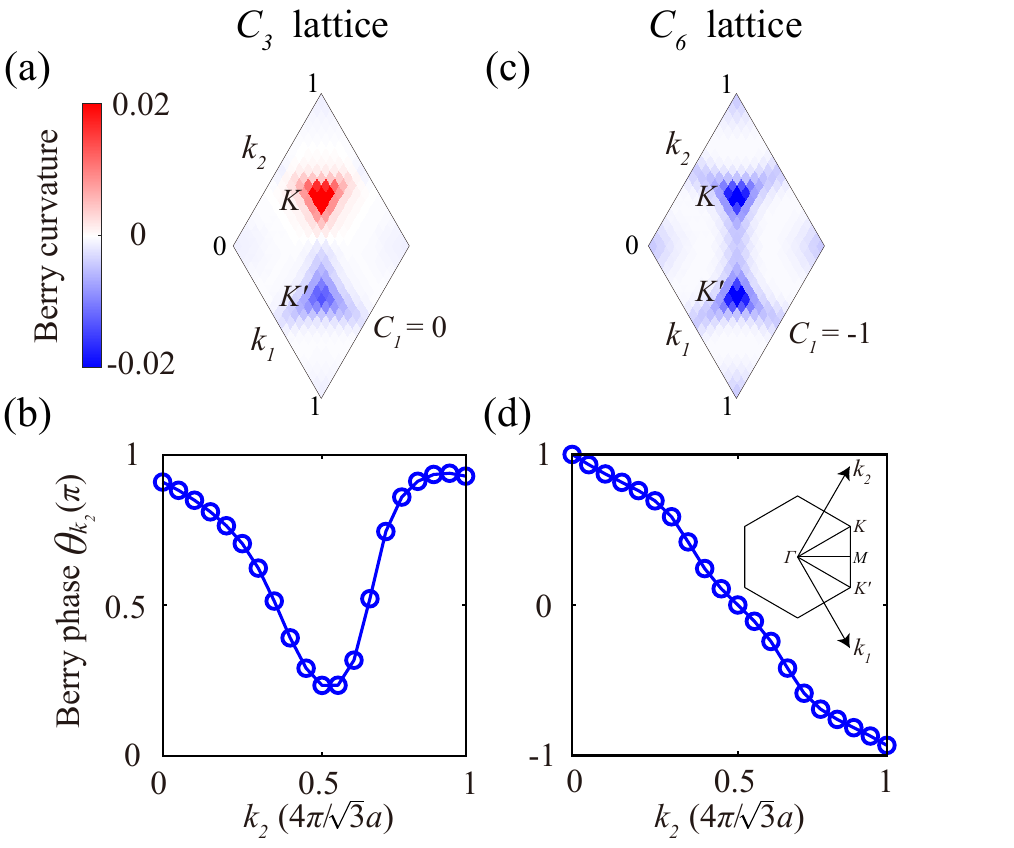} 
	   \caption{Topological invariants for the $C_{3}$ and $C_{6}$ lattices. (a), (c) Berry curvature distribution in the Brillouin zone, with corresponding Chern numbers for the lowest band of each lattice labeled alongside. (b), (d) Berry phase $\theta _{k_{2}} $ as a function of $k_{2}$ for the lowest band of the $C_{3}$ and $C_{6}$ lattices, respectively. The inset in (d) shows the Brillouin zone and corresponding reciprocal lattice basis vectors used in the calculations.}\label{fig:6}
\end{figure}  
\begin{equation}
    C_{n} = \frac{1}{2\pi } \int_{L_{i}} d\theta _{n,k_{i} } , 
	\end{equation}  
where $\theta _{n,k_{i}}$ is the Berry phase of the $n$th band, obtained by integrating the $j$-component of the Berry connection along the $k_{j}$ direction at a fixed $k_{i}$ ($k_{i} \ne k_{j}$)
	\begin{equation}
    \theta _{n,k_{i} } = \int_{L_{j}} dk_{j}{\textit{A}}_{n}^{j}\left ( \textbf{\textit{k}}\right ). 
	\end{equation} 
The Chern number associated with a band gap is the sum of the Chern numbers of all bands below the gap: $C_{gap} = \sum _{n} C_{n}$. 

Furthermore, in a $2$D system, the bulk polarization $P_{j}$ along direction $k_{j}$ can be expressed as \cite{PhysRevB.96.245115}
	\begin{equation}
    P_{j} = -\frac{1}{\left ( 2\pi  \right )^{2}}\int_{BZ}\mathrm{Tr}\left [\textit{A}^{j} \left ( \textbf{\textit{k}}\right )  \right ]d^{2}\textbf{\textit{k}} ,  
	\end{equation} 
where $\textit{A}^{j}\left ( \textbf{\textit{k}} \right )$ is the $j$-component of the Berry connection. Using the Wilson-loop approach, this expression can be further simplified into a more computationally tractable form \cite{PhysRevB.107.014105,Wang_2019}
 	\begin{equation}\label{B}
    P_{j} = \frac{1}{2\pi } \int_{L_{i}} d\theta _{n,k_{i} } . 
	\end{equation}  

In our calculations of the topological invariants for the $C_{3}$ and $C_{6}$ lattices, we focus on their first band gaps, each containing only a single band below the gap. Therefore, we set $n = 1$ in all topological invariant calculations. For the $C_{3}$ lattice, the regions of nonzero Berry curvature are localized near the high symmetry points $K$ and $K'$ in the Brillouin zone [see Fig. \ref{fig:6}(a)]. The Brillouin zone is represented as a parallelogram spanned by the reciprocal lattice basis vectors $k_{1}$ and $k_{2}$, as shown in the inset of Fig. \ref{fig:6}(d). As shown in Fig. \ref{fig:6}(a), the Berry curvature near $K$ and $K'$ exhibits opposite signs. Consequently, integrating the Berry curvature over the Brillouin zone yields a total Chern number of $0$. This behavior resembles the valley Hall effect \cite{PhysRevLett.116.093901,Lu2017,Wang_2019}. In contrast, for the $C_{6}$ lattice, the nonzero Berry curvature is also localized near the $K$ and $K' $ points [see Fig. \ref{fig:6}(c)]. However, it takes negative values at both high-symmetry points. Integrating over the Brillouin zone in this case yields a Chern number of $-1$. Additionally, for the $C_{3}$ lattice, we calculate the evolution of the Berry phase $\theta _{k_{2}} $ along the $k_{2}$ direction using the Wilson-loop approach, as shown in Fig. \ref{fig:6}(b). Using Eq. (\ref{B}), we obtain the bulk polarization $P_{1} = 1/3$. Similarly, the Berry phase $\theta _{k_{1}} $ evolution along the $k_{1}$ direction gives $P_{2} = 1/3$. For the $C_{6}$ lattice, the Wilson-loop calculation reveals a $2\pi$ winding of the Berry phase $\theta _{k_{2}} $ along the $k_{2}$ direction [Fig. \ref{fig:6}(d)], corresponding to a Chern number of $-1$, consistent with the result inferred from the Berry curvature distribution in Fig. \ref{fig:6}(c).

\bigskip

	\noindent{$^{*)}$zhanglei@sxu.edu.cn}\\
	\noindent{$^{\dagger)}$chenjun@sxu.edu.cn}
	\bibliography{myref1}
	
\end{document}